\documentclass[lettersize,journal]{IEEEtran}
\usepackage{amsmath,amsfonts}
\usepackage{algorithmic}
\usepackage{algorithm}
\usepackage{array}
\usepackage[caption=false,font=footnotesize]{subfig}
\usepackage{textcomp}
\usepackage{stfloats}
\usepackage{url}
\usepackage{verbatim}
\usepackage{graphicx}
\usepackage{cite}
\begin{document}

\title{Distribution-Aware Constellation Learning for Image Transmission}

\author{Xufeng Zhang, Yinhuan Huang, Jingkai Ying, Huan Liu, and Zhijin Qin,~\IEEEmembership{Senior Member,~IEEE}%
\thanks{Xufeng Zhang, Yinhuan Huang, Jingkai Ying, and Zhijin Qin are with the Department of Electronic Engineering, Tsinghua University, Beijing 100084, China, and also with the State Key Laboratory of Space Network and Communications, Beijing 100084, China (e-mail: xf-zhang22@mails.tsinghua.edu.cn; huangyh24@mails.tsinghua.edu.cn; yjk23@mails.tsinghua.edu.cn; qinzhijin@tsinghua.edu.cn).}%
\thanks{Huan Liu is with Shanghai Highgain Information Technology Co., Ltd., Shanghai 201700, China (e-mail: liuhuan@highgain.com.cn).}}

\maketitle

\begin{abstract}
Semantic communication has demonstrated significant potential for image transmission, especially in bandwidth-limited and low signal-to-noise ratio scenarios. However, most existing methods are based on analog transmission, which poses challenges to the compatibility with existing digital communication systems. Existing digital semantic communication methods commonly adopt conventional quadrature amplitude modulation constellations, which mismatch the empirical distribution of semantic features produced by the semantic encoder. This paper proposes a distribution-aware learnable modulation for semantic communication framework, which bridges semantic feature representations and discrete modulation through constellation learning. Specifically, a learnable constellation module, initialized with an amplitude phase shift keying geometric prior, is developed to refine the constellation geometry as a trainable codebook, enabling modulation symbols to better align with the distribution of semantic features. To enable end-to-end optimization, a two-stage training strategy is introduced, combining differentiable soft assignment with straight-through estimator. Simulation results show that the proposed framework consistently outperforms existing digital semantic communication schemes and achieves performance comparable to advanced analog methods.
\end{abstract}


\section{Introduction}
\IEEEPARstart{S}{e}mantic communication has recently emerged as a promising paradigm for efficient wireless transmission by focusing on task-relevant information rather than bit-level reconstruction\cite{10639525}. For wireless image transmission, conventional systems, which independently perform source coding, channel coding and modulation, may suffer from performance loss under finite blocklength and channel impairments. Joint semantic-channel coding (JSCC) provides an effective alternative by jointly optimizing semantic coding and channel coding to preserve task-relevant semantic information\cite{9398576}. 

The pioneering DeepJSCC\cite{8723589} showed that a neural transceiver can directly map images to channel symbols and outperform conventional digital schemes in bandwidth-limited or low signal-to-noise ratio (SNR) regimes. Building on advances in learned image compression\cite{balle2018variational}, recent works integrated entropy modeling into DeepJSCC, further improving rate-distortion performance\cite{9791398,10214392,11301684}. However, many JSCC frameworks rely on continuous channel symbols, limiting their compatibility with existing digital communication systems. To address the issue, digital JSCC has been introduced to map latent features onto finite symbol sets\cite{9838671}. Recent studies further developed joint semantic-channel coding and modulation scheme with dedicated differentiable modulation designs based on conventional quadrature amplitude modulation (QAM) constellations\cite{10495330,11276859,10778620}. 

Despite recent progress, a fundamental challenge lies in that most methods rely on fixed QAM constellations, while channel symbols produced by the JSCC encoder are often Gaussian-like and highly nonuniform\cite{yang2024digital}. This mismatch may lead to additional distortion when continuous semantic features are projected onto discrete modulation symbols. To address the issue, we develop a distribution-aware learnable modulation for semantic communication (DLM-SC). The proposed method introduces a learnable constellation codebook, where the constellation geometry is jointly optimized with the semantic transceiver. Amplitude phase shift keying (APSK) is adopted as a structured initialization prior for the codebook, providing a radial geometry suitable for nonuniform symbols.

However, DLM-SC still includes a non-differentiable operation, a differentiable training strategy is required for end-to-end optimization. A common solution is the straight-through estimator (STE)\cite{Bengio2013EstimatingOP}, which approximates the derivative of the quantization operator as a constant during backpropagation. However, in the early stage of training, many symbols are far from the constellation points, causing abrupt output jumps in the forward pass and unstable gradient propagation in backpropagation. To address the difficulty, we further develop a two-stage training strategy that combines differentiable soft assignment with the STE method. The main contributions of this letter are summarized as follows:

\begin{itemize}
\item We propose a distribution-aware digital semantic communication framework by representing the modulation constellation as a trainable codebook, which enables the constellation geometry to adapt to the empirical distribution of semantic features produced by the JSCC encoder.

\item We develop a two-stage training strategy for the modulation non-differentiability. Specifically, a differentiable soft assignment mechanism is first employed to stabilize optimization, followed by STE-based fine-tuning.

\item Extensive experiments demonstrate that the proposed DLM-SC consistently outperforms existing digital semantic communication schemes and achieves competitive performance relative to advanced analog methods.
\end{itemize}

The rest of this letter is organized as follows. Section II presents the proposed DLM-SC framework. Section III describes the proposed method. Section IV provides the simulation results. Finally, Section V concludes this letter. 

\section{System Overview}

\begin{figure}[t!]
\centerline{\includegraphics[width=0.48\textwidth]{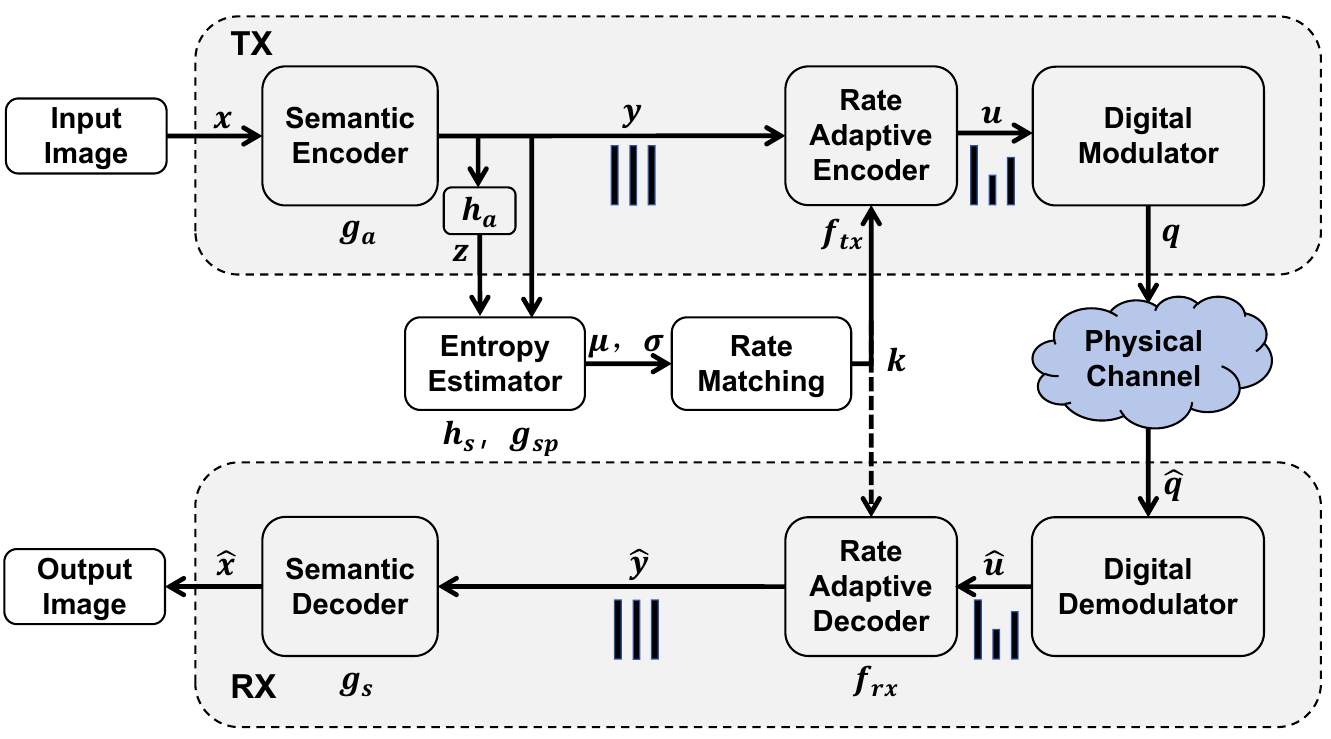}}
\caption{The overall architecture of distribution-aware learnable modulation for semantic communication.}
\label{fig_overview}
\end{figure}

Based on the image semantic communication system with quadtree partition-based coding\cite{11301684}, we propose DLM-SC. The overall architecture of the proposed communication framework is illustrated in Fig. \ref{fig_overview}. 

Given an input image $\mathbf{x}\in \mathbb{R}^{H \times W \times 3}$, the semantic encoder first extracts the latent representation $\mathbf{y}$ as

\begin{equation}
\mathbf{y}=g_a(\mathbf{x}),\label{eq:semantic_encoder}
\end{equation}
where $g_a(\cdot)$ denotes the semantic encoder and $\mathbf{y}\in\mathbb{R}^{H_y \times W_y \times C_y}$ is the extracted semantic feature tensor. To characterize the distribution of $\mathbf{y}$, an entropy estimator is introduced. Specifically, a hyper-prior encoder $h_a(\cdot)$ first produces an auxiliary latent $\mathbf{z}$ as

\begin{equation}
\mathbf{z}=h_a(\mathbf{y}),\label{eq_z}
\end{equation}
whose quantized version $\mathbf{\hat{z}}=Q(\mathbf{z})$ is decoded by the hyper-prior decoder $h_s(\cdot)$ to provide global prior information. Then, a quadtree partition-based space-channel context model progressively estimates the conditional statistics of the latent partitions. For the $i$-th partition, the corresponding entropy parameters are written as

\begin{equation}
(\boldsymbol{\mu}_i, \boldsymbol{\sigma}_i^2) = g_{\text{sp}}^{(i)} \big( \hat{\mathbf{y}}_{\text{ctx}}^{(i)}, h_s(\mathbf{\hat{z}}) \big),\label{eq_prior}
\end{equation}
where $\hat{\mathbf{y}}_{\text{ctx}}^{(i)}$  is the previously reconstructed quadtree context and $g_{\text{sp}}^{(i)}(\cdot)$ is the parameter predictor. The entropy estimates are then used for rate matching. For each spatial unit, the symbol-length factor is computed as

\begin{equation}
k_{m,n} = Q_k \left( \sum_{c=1}^{C_y} -\eta \log_2 p\left(y_{m,n,c} \,\big|\, \hat{\mathbf{z}},\hat{\mathbf{y}}_{\text{ctx}}\right) \right),\label{eq_k}
\end{equation}
where $\eta$ controls the transmission rate and $Q_k(\cdot)$ maps the entropy accumulation to a predefined discrete rate set. Here, $m$ and $n$ denote the spatial indices of the semantic feature tensor $\mathbf{y}$, and $c$ denotes the channel index. The collection of $\{k_{m,n}\}$ forms a rate map $\mathbf{k}$, which is transmitted as side information to assist receiver-side reconstruction. 

Guided by $\mathbf{k}$, the rate-adaptive encoder maps each spatial unit to a variable-length real-valued transmission vector as

\begin{equation}
\mathbf{u}_{m,n} = f_{\text{tx},k_{m,n}}\!\left(\mathbf{y}_{\text{sp}}^{(m,n)}\right) \odot \mathbf{m}_{k_{m,n}},\label{eq_rate}
\end{equation}
where $f_{\text{tx},k_{m,n}}(\cdot)$ denotes the rate-dependent feature mapping, $y_{sp}^{(m,n)} = y[m,n,:] \in \mathbb{R}^{C_y}$ denotes the semantic feature vector at spatial location $(m,n)$ and $\mathbf{m}_{k_{m,n}}= [1, \dots, 1, 0, \dots, 0]$ is a binary mask that retains the first $k_{m,n}$ active dimensions. After masking, the retained entries from all spatial units are concatenated and flattened into a one-dimensional sequence $\mathbf{u} = [u_0, \dots, u_{2t}, u_{2t+1}, \dots, u_{L_u-1}]$, where $L_u$ is the total number of active entries. The flattened sequence is then grouped into complex symbols through I/Q pairing as

\begin{equation}
q_t = u_{2t} + ju_{2t+1}, \quad q_t \in \mathbb{C},\label{eq_q}
\end{equation}
zero padding is applied when $k_{m,n}$ is odd. We use the channel bandwidth ratio (CBR) to measure the actual channel usage\cite{8723589}. The complex symbols are fed into a learnable modulation module and mapped to discrete channel symbols as

\begin{equation}
s_t = \mathcal{M}\left(q_t\right),\label{eq_s}
\end{equation}
where $\mathcal{M}(\cdot)$ denotes the proposed modulator. The detailed design is presented in Section III. The modulated symbols are then transmitted through the wireless channel. For an additive white Gaussian noise (AWGN) channel, the received signal is expressed as 

\begin{equation}
r_t=s_t+n_t,\label{eq_awgn}
\end{equation}
where $n_t$ is the additive complex Gaussian noise. At the receiver, the received symbols are demodulated to recover the constellation points, whose I/Q components are rearranged into the real-valued representation $\hat{\mathbf{u}}$. With the side information $\mathbf{k}$, the rate-adaptive feature decoder progressively reconstructs the latent representation as

\begin{equation}
\hat{\mathbf{y}} = f_{\text{rx}}\left(\hat{\mathbf{u}},\mathbf{k}\right),\label{eq_reconstruction_y}
\end{equation}
where $f_{rx}(\cdot)$ denotes the rate-adaptive feature decoder. Finally, the semantic decoder reconstructs the image as 

\begin{equation}
\mathbf{\hat{x}}=g_s(\mathbf{\hat{y}}),\label{eq_reconstruction_x}
\end{equation}
where $g_s(\cdot)$ is the semantic decoder. Through the above procedure, the proposed framework establishes an end-to-end digital semantic transmission pipeline. 

\begin{figure}[t!]
\centerline{\includegraphics[width=0.47\textwidth]{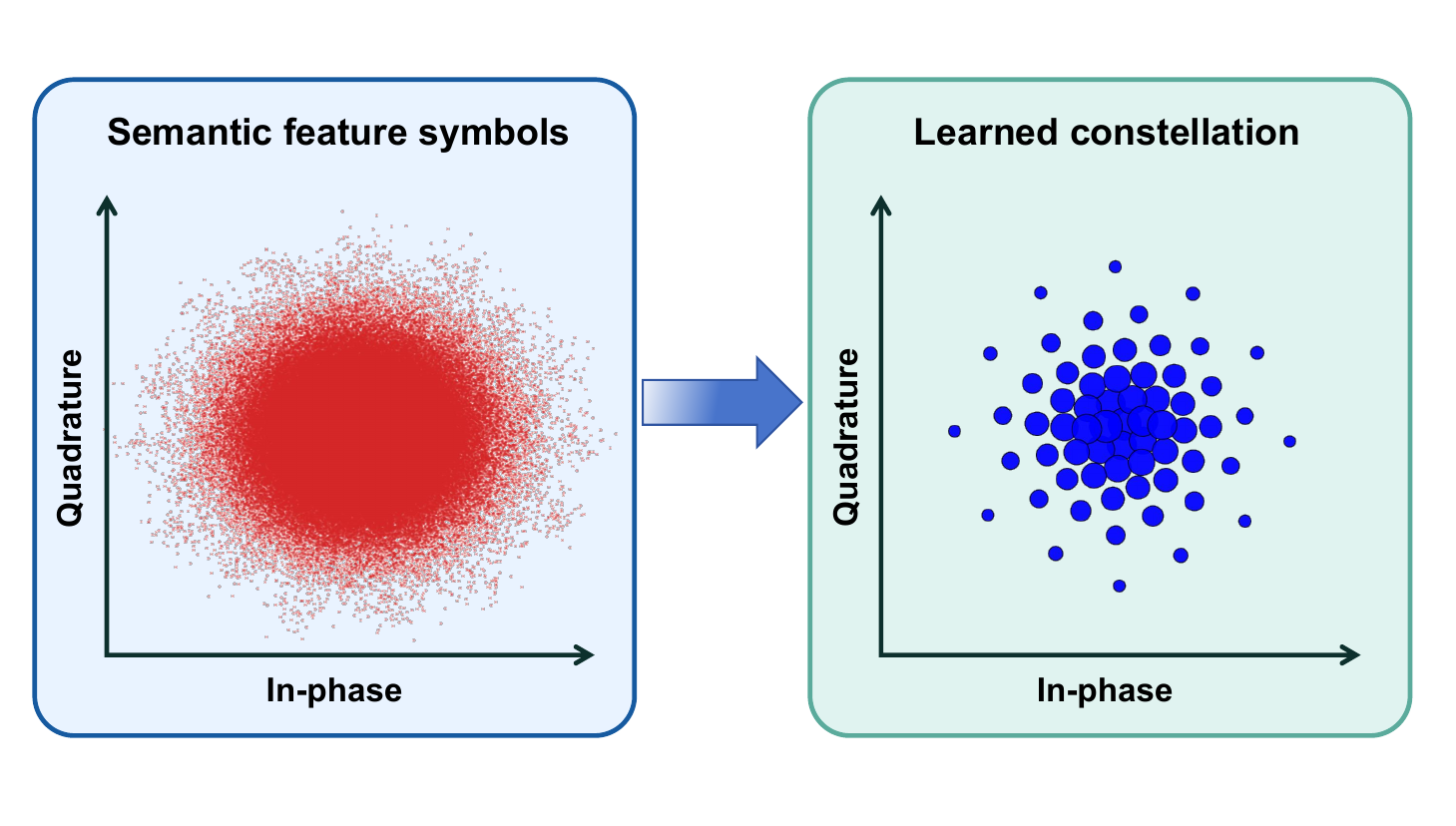}}
\caption{Constellation distribution before and after learnable modulation, larger markers indicate higher selection probabilities of the constellation points.}
\label{fig_before_after_modu}
\end{figure}

\section{Proposed Method }
This section presents the proposed distribution-aware learnable modulator and its two-stage training strategy.

\subsection{Learnable Constellation Module  }

As shown in Fig. \ref{fig_before_after_modu}, the encoder outputs highly nonuniform semantic symbols that roughly follow a Gaussian-like distribution, whereas fixed QAM adopts a Cartesian geometry. Such a mismatch may cause distortion when continuous semantic features are projected onto discrete digital symbols. 

To alleviate this issue, we introduce a distribution-aware learnable constellation module, which enables the discrete modulation space to adapt to the empirical distribution of compressed semantic features. In this work, we adopt an $M$-ary APSK structure as the default initialization, which provides a geometry-aware prior rather than an unconstrained initialization. The initial constellation is constructed as 

\begin{equation}
\mathcal{C}_M = \bigcup_{l=1}^{L} \left\{ r_l e^{j\theta_{l,n}} \mid n = 1, \dots, N_l \right\}, \quad \sum_{l=1}^{L} N_l = M,\label{eq_CM}
\end{equation}
with
\begin{equation}
\theta_{l,n} = \frac{2\pi n}{N_l} + \phi_l,  \label{eq_theta}
\end{equation}
where $L$ denotes the number of rings, $N_l$ is the number of points on the $l$-th ring, $r_l$ is the corresponding radius, and $\theta_{l,n}$ is the angular position of the $n$-th point,  $\phi_l$ is the phase offset, the ring radii are initialized according to the quantiles of a Rayleigh distribution\cite{6583813}. Let $m_1 = \log_2 L$ denote the number of amplitude bits. The unnormalized radius of the $l$-th ring is computed as 

\begin{equation}
r_l = \sqrt{-\ln\left(1 - \left(l - \frac{1}{2}\right) 2^{-m_1}\right)}, \quad l = 1, \dots, L. \label{eq_rl}
\end{equation}

\begin{figure*}[t!]
    \centering
    \includegraphics[width=0.9\textwidth]{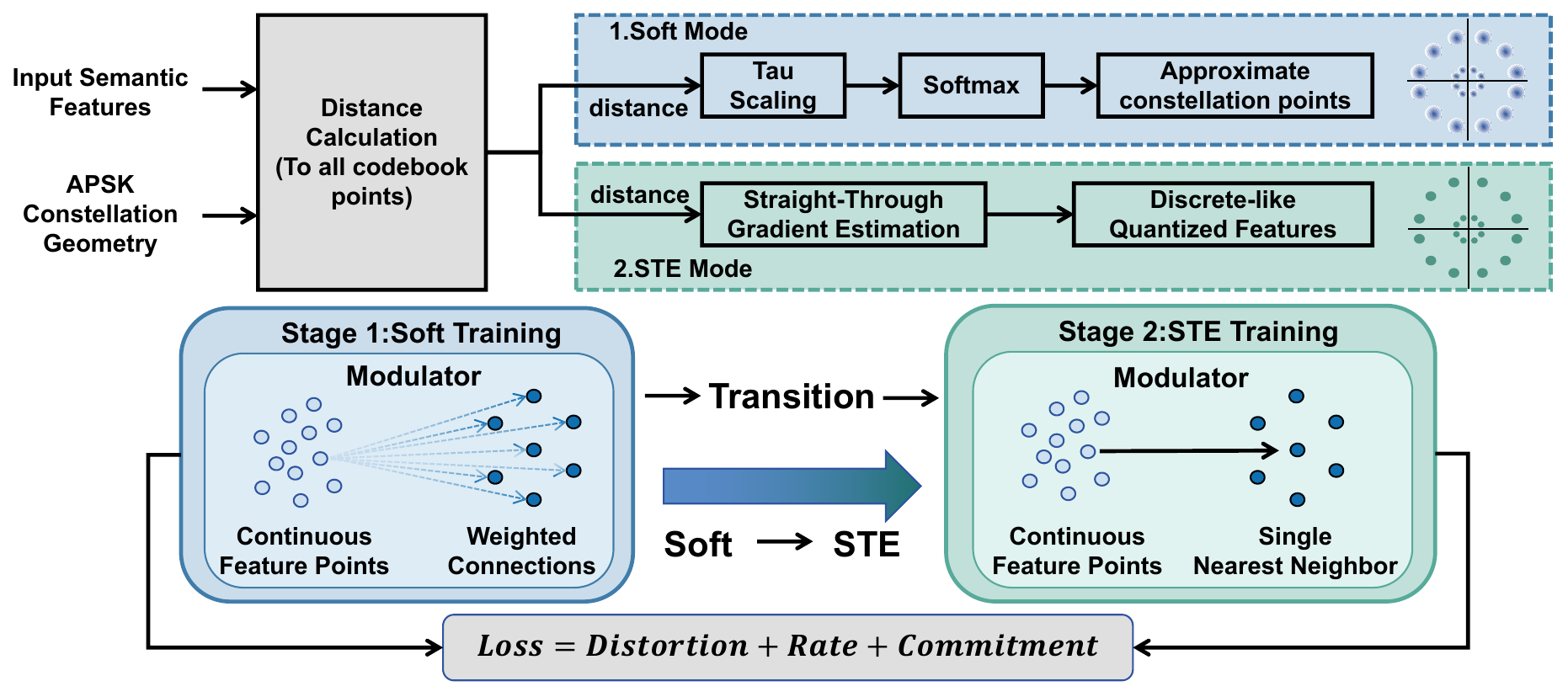}
    \caption{Illustration of the two-stage training strategy for the proposed modulation module. The upper part shows the unified modulation framework. The lower part shows the specific training process and training objectives.}
    \label{fig_training}
\end{figure*} 

This construction allocates the rings according to the radial statistics of a two-dimensional Gaussian source, which is consistent with the approximately Gaussian-like distribution of semantic channel symbols. The radii are then normalized to satisfy the unit average-power constraint

\begin{equation}
\bar{r}_l = \frac{r_l}{\sqrt{\frac{1}{M} \sum_{i=1}^L N_i r_i^2}}. \label{eq_rl_bar}
\end{equation}

After initialization, the constellation is refined as a learnable codebook $\mathcal{C}_M = \left\{ c_m \right\}_{m=1}^{M}$, where the radius and phase of each constellation point are updated during end-to-end training, allowing the geometry to adapt to the empirical feature distribution. To satisfy the average transmit-power constraint, each constellation point is normalized as 

\begin{equation}
\bar{c}_m = \sqrt{\frac{1}{\frac{1}{M} \sum_{j=1}^M |c_j|^2}} \, c_m. \label{eq_cm}
\end{equation}

The digital modulation is realized by nearest-neighbor mapping from the continuous symbol $q_{in}$ to the normalized constellation as

\begin{equation}
s_t = q_{out}^{\text{hard}}= \arg\min_{\bar{c}_m \in \mathcal{C}_M} \left\| q_{in} - \bar{c}_m \right\|^2 ,\label{eq_hard}
\end{equation}
where $s_t$ is the transmitted discrete channel symbol. The distribution of semantic features and the constellation diagram are shown in Fig. \ref{fig_before_after_modu}. The learned constellation shows a Gaussian-like pattern. This confirms that the proposed learnable constellation adapts to the empirical distribution of semantic symbols. Since modulation only requires a nearest-neighbor codebook lookup, the additional complexity is negligible.

\subsection{Two-Stage Training Strategy }
 To address the challenge of nearest-neighbor mapping, we adopt a two-stage training strategy as shown in Fig. \ref{fig_training}.

In stage I, we employ a differentiable soft assignment to approximate the hard quantization\cite{9838671}. For a given symbol $q_{in}$, the distance between it and each constellation point is

\begin{equation}
d_m(q_{in}) = \left\| q_{in} - \bar{c}_m \right\|_2. \label{eq_distance}
\end{equation}

The soft assignment weights are given by

\begin{equation}
w_m(q_{in}) = \frac{\exp\left(-d_m(q_{in})/\tau\right)}{\sum_{j=1}^{M} \exp\left(-d_j(q_{in})/\tau\right)}. \label{eq_wm}
\end{equation}

Based on the weights, the relaxed modulated symbol is obtained as a weighted combination of all constellation points:

\begin{equation}
q_{out} = \sum_{m=1}^{M} w_m(q_{in})\, \bar{c}_m, \label{eq_stage1}
\end{equation}
where $\tau$ denotes the temperature parameter.  As $\tau$ gradually decreases, soft assignment asymptotically approaches hard decision-making, guiding symbols toward the learned constellation while stabilizing early optimization. Therefore, this stage can encourage the semantic symbols and the constellation points to move closer to each other while avoiding severe optimization instability in early training. 

\begin{figure*}[t!]
    \centering
    \includegraphics[width=0.94\textwidth]{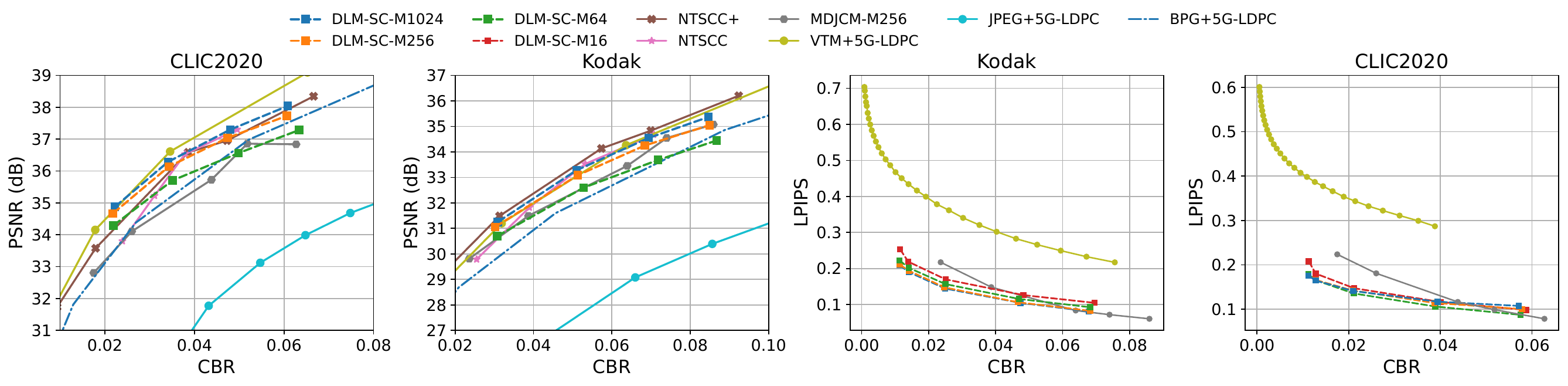}
    \caption{Performance comparison of different methods under varying CBR conditions. Lower LPIPS and higher PSNR indicate better quality.}
    \label{fig_PSNR_LPIPS}
\end{figure*}

In stage II, we perform fine-tuning using the STE method with the learnable codebook fixed. The hard decision is defined as (16), the STE formulation is

\begin{equation}
q_{out} = q_{in} + \text{sg}\!\left(q_{out}^{\text{hard}} - q_{in}\right), \label{eq_stage2}
\end{equation}
where $\text{sg}(\cdot)$ represents the stop-gradient operator. This formulation makes the forward pass consistent with practical hard-decision modulation, while preserving gradient propagation. During the training, the overall objective is defined as 

\begin{equation}
\mathcal{L} = \lambda D(\mathbf{x}, \hat{\mathbf{x}}) + \beta R + \gamma \mathcal{L}_{\text{commit}}, \label{eq_loss}
\end{equation}
with
\begin{equation}
\mathcal{L}_{\text{commit}} = \mathbb{E}\left[\left\| q_{in} - \text{sg}(q_{out}^{\text{hard}}) \right\|_2^2\right], \label{eq_losscommit}
\end{equation}
where $D(\cdot)$ denotes the reconstruction distortion (MSE), $R$ is the rate proxy derived from entropy modeling, $\mathcal{L}_{\text{commit}}$ is used to constrain the continuous representations to stay close to their corresponding hard constellation points\cite{esser2021taming}. The coefficients $\lambda$, $\beta$, and $\gamma$ are weighting factors used to balance the distortion, rate, and commitment loss terms, respectively.

\section{Simulation Results}

\subsection{Simulation Settings}

The model is trained on the 8000 largest images from the ImageNet dataset\cite{5206848}. During training, each image is randomly cropped to $256\times256$. We employ the Adam optimizer for parameter updates. The modulation order is selected from (4, 16, 64, 256, 1024). We train the model at an SNR of 10 dB. In stage I, we employ the soft relaxation approach, with a learning rate of $1\times10^{-4}$, and the temperature $\tau$ is exponentially annealed once per epoch as $\tau^{(e)}=\max(0.05,\,0.25\cdot0.97^{e})$.  In stage II, we perform the fine-tuning with a learning rate of $1\times10^{-5}$ and the temperature annealing is disabled. The loss weights are set to $\beta=0.05$ and $\gamma=50$. Different CBR operating points are obtained by training models with different $\lambda$ values. We investigate the performance of the proposed method over AWGN channels. 

We evaluate our method on different datasets. The Kodak dataset\cite{KodakDataset} consists of 24 high-quality images with a resolution of $768\times512$. We also consider the CLIC2020 dataset\cite{CLIC2020}, whose images have resolutions ranging from $1K$ to $2K$. For comparison, we consider both analog and digital baselines. Firstly, we include NTSCC\cite{9791398} and its variant NTSCC+\cite{10214392}. In addition, we also introduce representative digital methods, including MDJCM\cite{10778620}. For conventional methods, we combine the advanced image codec BPG/JPEG/VTM with the channel codec LDPC. The coding rate and modulation order are selected according to Table 5.1.3.1-1 for the Physical Downlink Shared Channel (PDSCH) in 3GPP TS 38.214\cite{3gpp38214}. We sweep the modulation and coding scheme (MCS) indices and report the best performance among them.

\subsection{Performance Analysis}

\begin{figure}[t!]
\centerline{\includegraphics[scale=0.36]{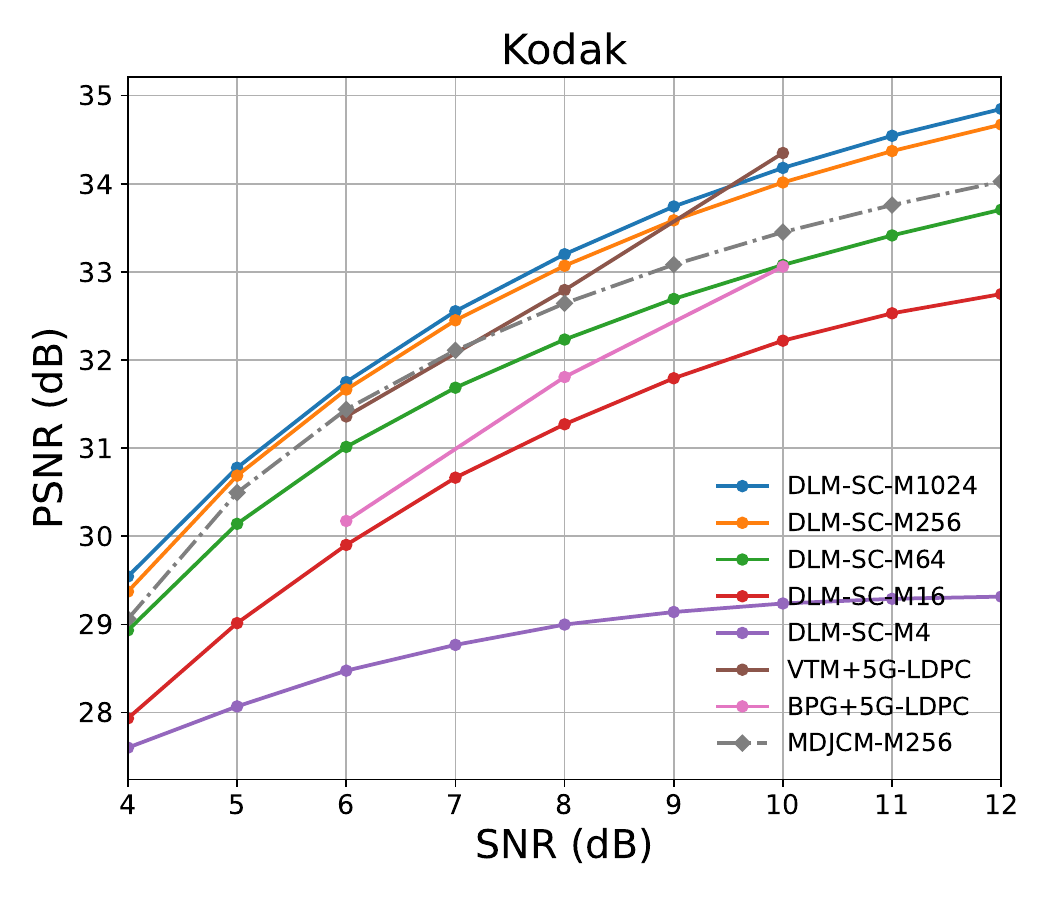}}
\caption{The PSNR performance versus SNR over Kodak dataset.}
\label{fig_SNR}
\end{figure}

Fig. \ref{fig_PSNR_LPIPS} reports Peak Signal-to-Noise Ratio (PSNR) and Learned Perceptual Image Patch Similarity (LPIPS) versus CBR over the AWGN channel at 10 dB. We consider different modulation orders, where the model is trained with a fixed modulation order. For PSNR, the results show that DLM-SC outperforms other digital modulation models and approaches the performance of advanced analog JSCC models. For LPIPS, our proposed method outperforms the benchmark models in terms of perceptual quality, and approaches the advanced MDJCM model at high CBR levels. Furthermore, the proposed method achieves more pronounced performance gains at lower CBR levels. This is because, at low CBR, each transmitted symbol carries more reconstruction information. Therefore, a mismatch between discrete modulation and feature distribution leads to performance degradation. Fig. \ref{fig_SNR} depicts the PSNR performance as SNR varies. All transmission schemes are tested under a constant CBR value of 0.0625. The results show that the proposed model maintains its performance advantage in different SNR conditions, especially in the low-SNR regime, where it clearly outperforms conventional schemes and demonstrates stronger robustness to channel noise.

\begin{figure}[t!]
\centerline{\includegraphics[scale=0.36]{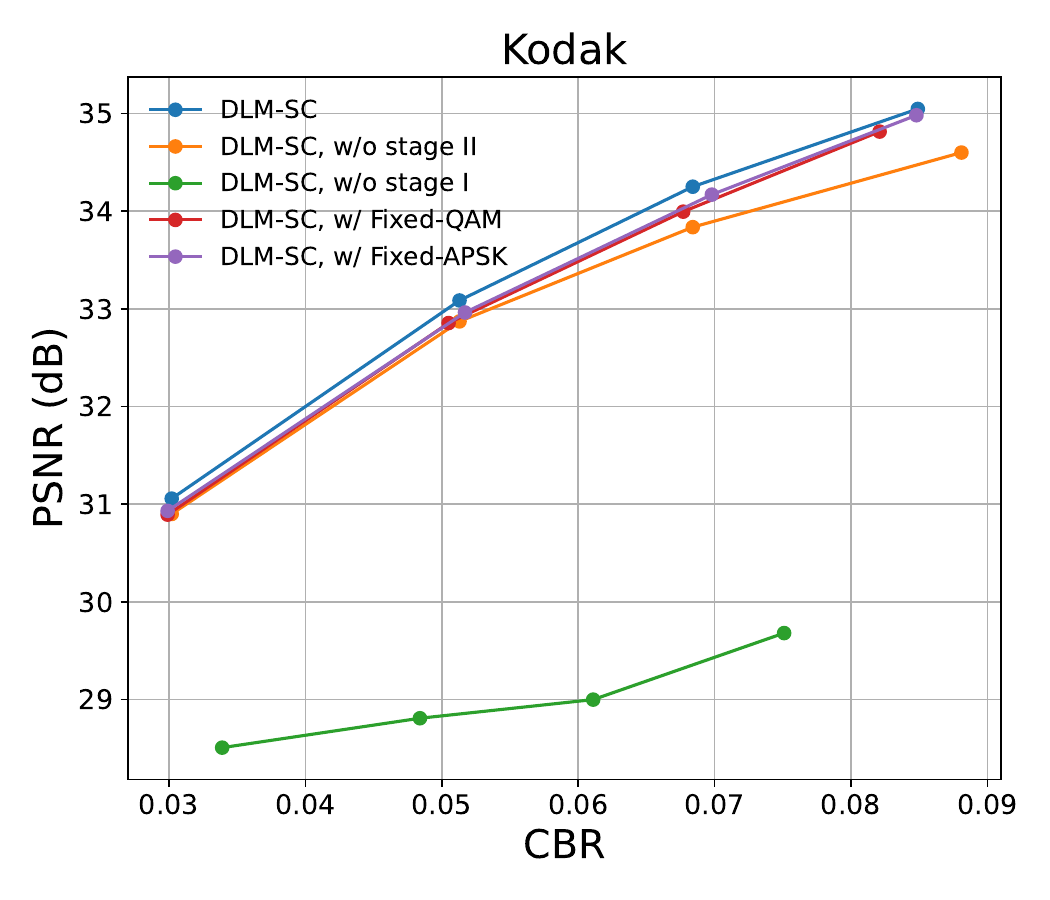}}
\caption{Ablation results of the proposed DLM-SC on the Kodak dataset.}
\label{fig_ablation}
\end{figure}

For ablation studies, we ablate DLM-SC by removing key components, verifying the effects of the learnable codebook and the two-stage training strategy, respectively. For the sake of fairness, the same number of training epochs is adopted in all four experiments. In Fig. \ref{fig_ablation}, ``DLM-SC, w/o stage II'' denotes the variant trained only with stage I, while ``DLM-SC, w/o stage I'' denotes the variant trained only with stage II. The experiment shows that when only the STE method is used for training, the optimization process becomes difficult to converge. Removing stage II leads to a smaller but consistent performance loss, demonstrating that the fine-tuning phase is still necessary to better match practical discrete modulation. In addition, Fixed-APSK slightly outperforms Fixed-QAM, suggesting that the radial APSK geometry better matches the distribution of semantic symbols than the QAM structure. Both fixed constellations are consistently inferior to DLM-SC, demonstrating the importance of adaptively learning the constellation according to the semantic feature distribution.

\section{Conclusion}

In this letter, we propose DLM-SC, a digital semantic image transmission framework with a distribution-aware learnable constellation and a two-stage training strategy. The experiments show that DLM-SC outperforms existing methods in various performance metrics, especially under low CBR conditions. The results demonstrate the effectiveness of distribution-aware constellation learning for digital semantic transmission.

\bibliographystyle{IEEEtran}
\bibliography{IEEEabrv,reference}

\end{document}